\documentclass[prl,aps,twocolumn, showpacs]{revtex4}

\usepackage{graphicx}
\usepackage{amsmath}
\usepackage{bm}
\usepackage[latin1]{inputenc}
\usepackage{amstext}
\usepackage{amsxtra}
\usepackage{times}
\newcommand{\bs}{\boldsymbol}
\newcommand{\beq}{\begin{equation}}
\newcommand{\eeq}{\end{equation}}
\newcommand{\bpm}{\begin{pmatrix}}
\newcommand{\epm}{\end{pmatrix}}
\newcommand{\bea}{\begin{eqnarray}}
\newcommand{\eea}{\end{eqnarray}}

\newcommand{\kB}{k_{\rm B}}
\newcommand{\ldB}{\lambda_{\rm T}}
\newcommand{\lop}{\lambda_{\rm opt}}

\newcommand{\Ep}{E_{\rm p}}
\newcommand{\Ek}{E_{\rm k}}
\newcommand{\Ei}{E_{\rm i}}

\newcommand{\Is}{I_{\rm sat}}
\newcommand{\If}{I_{\rm f}}
\newcommand{\Ii}{I_{\rm i}}
\newcommand{\ntD}{n^{\rm (2D)}}

\newcommand{\ntDh}{n^{\rm (2D)}_{\rm hom}}

\newcommand{\TOF}{{\sc T}o{\sc F}}

\newcommand{\EOS}{{\sc EoS}}

\begin{document}
\title{Exploring the thermodynamics of a two-dimensional Bose gas}
\date{\today}
\author{Tarik Yefsah, R\'{e}mi Desbuquois, Lauriane Chomaz, Kenneth J. G\"{u}nter and Jean Dalibard}
\affiliation{Laboratoire Kastler Brossel, CNRS, UPMC, Ecole Normale Sup\'erieure, 24 rue Lhomond, F-75005 Paris, France}

\begin{abstract}
Using \emph{in situ} measurements  on a quasi two-dimensional, harmonically trapped $^{87}$Rb gas, we infer various equations of state for the equivalent homogeneous fluid. From the dependence of the total atom number and the central density of our clouds with the chemical potential and temperature, we obtain the equations of state for the pressure and the phase-space density.  Then using the approximate scale invariance of this two-dimensional system, we determine the entropy per particle. We measure values as low as $0.06\,\kB$ in the strongly degenerate regime, which shows that a 2D Bose gas can constitute an efficient coolant for other quantum fluids. We also explain how to disentangle the various contributions (kinetic, potential, interaction) to the energy of the trapped gas using a time-of-flight method, from which we infer the reduction of density fluctuations in a non fully coherent cloud.
\end{abstract}

\maketitle

Physical properties of homogeneous matter at thermal equilibrium are characterized by an equation of state (\EOS), \emph{i.e.} a relationship between some relevant state variables. For a fluid of particles, possible \EOS's consist in expressions of  pressure, density or entropy as functions of  temperature $T$ and chemical potential $\mu$. For an ideal gas the \EOS\ can be calculated exactly  in any dimension for a Bose or Fermi gas. In the presence of interactions, one has to resort to approximations or numerical calculations, and a comparison with experiments is crucial to test their validity. Trapped atomic gases at thermal equilibrium provide a powerful tool for this purpose \cite{Ho:2009}. Within the local density approximation, any intensive state variable takes at a point $\bs r$ in the trap the same value as in a homogenous system with the same temperature and the shifted chemical potential $\mu-V(\bs r)$, where $V(\bs r)$ is the confining potential.

The case of an interacting two-dimensional (2D) Bose fluid is particularly interesting in this context. Firstly, at non-zero temperature the Mermin--Wagner theorem precludes  Bose--Einstein condensation \cite{Mermin:1966,Hohenberg:1967}. Therefore the  \EOS\ is expected to be continuous at any point, in spite of the existence of a superfluid, infinite-order phase transition, which is of the Berezinskii--Kosterlitz--Thouless (BKT) type \cite{Berezinskii:1971,Kosterlitz:1973}.  Secondly the \EOS\ of a 2D Bose fluid is scale invariant \cite{Prokofev:2002} in the so-called quasi--2D regime. The later refers to the experimentally relevant situation where single-particle motion is frozen along one axis, making it thermodynamically 2D, but where collisions still keep their 3D character \cite{Petrov:2000a}. The (approximate) scale invariance originates from the fact that in this regime, the coupling strength $\tilde g$ is an energy-independent dimensionless coefficient, and thus provides no energy, nor length scale, in contrast with the 1D or 3D cases. This implies in particular that dimensionless thermodynamic variables such as the phase space density ${\cal D}$ or the entropy per particle ${\cal S}$ are functions of the ratio $\mu/\kB T$ only, with $\tilde g$ as a parameter.

Recent experiments with trapped 2D Bose atomic gases have demonstrated the existence of a BKT-type transition at a threshold phase space density. One line of investigation exploited matter-wave interference to monitor the appearance of an extended coherence  in the sample  \cite{Hadzibabic:2006,Clade:2009}, and another approach used a time-of-flight (\TOF) technique to measure the momentum distribution of the gas \cite{Tung:2010}. The steady-state scale invariance was verified in \cite{Hung:2011}. In this paper we present a detailed experimental investigation of several thermodynamic properties of a 2D Bose gas. We describe measurements of the \EOS\ for the pressure from a count of the total atom number of a trapped 2D Bose gas, with a wide range of thermodynamic parameters. From the same set of data we use the central spatial density to access the \EOS\ for phase space density. Combining these two measured \EOS\ with the scale invariance we obtain the \EOS\ for the entropy per particle. We show that this quantity rapidly decreases around the superfluid transition point and then approaches zero in the highly degenerate regime. We also present an original method to extract from a \TOF\ in only one direction of space, the various contributions (kinetic, potential, interaction) to the total energy of the trapped gas. This method is applicable to any low-dimensional fluid. Here it shows that density fluctuations of our 2D Bose gas are essentially frozen even when its thermal, non coherent fraction is significant.

Our 2D Bose gases are prepared along the lines detailed in \cite{Rath:2010b}. We start with a 3D Bose--Einstein condensate of $^{87}$Rb atoms confined in a magnetic trap in their $F=m_F=2$ internal ground state with an adjustable temperature. We slice a horizontal sheet of atoms with an off-resonant, blue-detuned laser beam with an intensity node in the plane $z=0$. It provides a strong confinement along the direction perpendicular to this plane, with oscillation frequency  $\omega_z/2\pi=1.9\,(2)$\,kHz, which correspond to the interaction strength $\tilde g=\sqrt{8\pi}a/\ell_z \approx 0.1$, where $a$ is the 3D scattering length and $\ell_z=\sqrt{\hbar/m\omega_z}$. The  energy  $\hbar \omega_z$ is similar to or larger than the thermal energy $\kB T$ and the interaction energy per particle, so that most of the atoms occupy the ground state of the vibrational motion along $z$. The magnetic trap provides a quasi-isotropic confinement in the $xy$ plane with  frequency $\omega\,/\,2\pi=20.6\,(1)$\,Hz \cite{footnote1}.

\begin{figure}[tbp]
\includegraphics{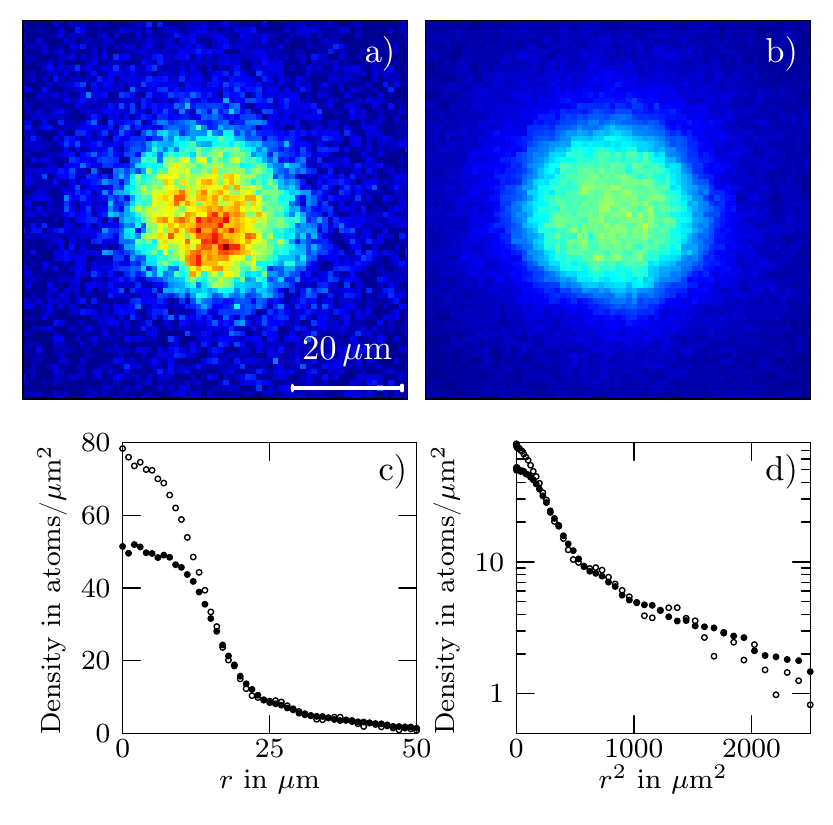}
\vspace{-0.5cm}
\caption{(Color on line) Absorption imaging of quasi-2D clouds of $^{87}$Rb atoms.
(a) Image obtained with a short pulse ($\sim2\mu$s) of an intense probe beam ($I/\Is = 40$). (b) Image obtained with a longer pulse (50 $\mu$s) of a weak probe beam ($I/\Is = 0.5$). (c) and (d) Radial density profiles for image (a) (hollow circles $\circ$) and image (b) (filled circles $\bullet$) in linear (c) and logarithmic (d) scales.}
\label{Fig:Cloud_images}
\end{figure}

After an equilibration time of $3$ seconds in the combined magnetic+laser trap, we measure the \emph{in situ} density distribution of the gas by performing absorption imaging with a probe beam propagating along the vertical axis. The conventional procedure where one uses a weak probe beam with an intensity $I$ well below the saturation intensity $\Is$ is problematic in this context \cite{Rath:2010b}. Indeed for the relevant range of temperatures (40--150~nK), the atomic thermal wavelength $\ldB$ is comparable to the optical wavelength used for probing, $\lop=780$~nm. Consequently in the highly degenerate region of the gas (${\cal D}\equiv \ntD\ldB^2 \gg 1$), the average distance between neighboring atoms is much smaller than $\lop$ and the absorption of a weak probe is strongly perturbed by collective effects. To circumvent this problem we probe the gas with a short pulse (duration $\sim 2\,\mu$s) of an intense probe beam (typically $I/\Is=40$ to 100) \cite{Reinaudi:2007}. The interaction of any given atom with light is then nearly independent of its neighbors.

High-intensity imaging, which was also used in \cite{Hung:2011}, provides a faithful measurement of the atomic distribution in the central region of the trap, where the density is large. However the quality of the images suffers from a large photon shot noise, which spoils the detection of the low-density regions of the cloud (Fig.~\ref{Fig:Cloud_images}a). In order to probe reliably these regions on which we base our determination of temperature and chemical potential, we complement the high-intensity imaging procedure by the conventional low-intensity one (Fig.~\ref{Fig:Cloud_images}b).  In practice for any set of parameters to be studied, we perform one run of the experiment with high-intensity imaging, and one with low-intensity imaging immediately after. The reproductibility of the experiment is checked by acquiring several pairs of images for a given set of experimental parameters.

The procedure for image processing is detailed in the Auxiliary Material. In short, for each pair of images it provides the temperature $T$, the chemical potential at center $\mu$ and the density $n(\bs r)$ at any pixel of the image. We assume the atoms in the excited states of the $z$ motion to be described by the Hartree--Fock mean-field (HFMF) theory \cite{Hadzibabic:2008,Bisset:2009,Holzmann:2010,Tung:2010}; therefore, we can self-consistently calculate the populations of the excited states, and subtract it from $n(\bs r)$
in order to obtain the density distribution $n_0(\bs r)$ in the ground state. The validity of this procedure was checked by analyzing  the results of a  quantum Monte Carlo calculation for a range of parameters similar to ours \cite{Holzmann:2008a}.

\begin{figure*}[tbp]
\begin{center}
\includegraphics{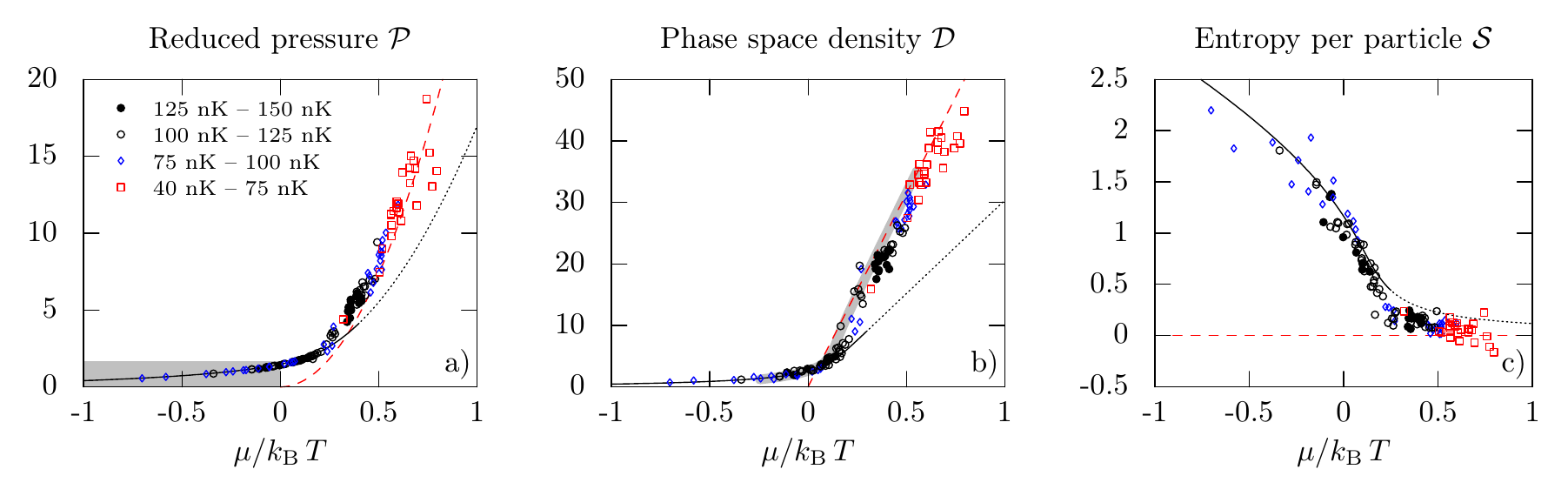}
\end{center}
\vspace{-1cm}
\caption{(Color on line) Equations of state for (a) the reduced pressure ${\cal P}$, (b) the phase space density ${\cal D}$ and (c) the entropy per particle ${\cal S}$. The Hartree-Fock mean field prediction is plotted in full line and extended in dotted line beyond the expected superfluid transition. The dashed line indicates the Thomas-Fermi prediction. In (a) the grey area indicates the region of parameter space accessible to an ideal gas. In (b) the thick grey line indicates the prediction from \cite{Prokofev:2002}.}
\label{Fig:EoS}
\end{figure*}

We start our thermodynamic analysis by inferring the pressure $P(\mu, T)$ of the homogeneous gas from our measurements. Here we adapt to the two-dimensional case the technique presented in \cite{Ho:2009} and that has successfully been used in 3D for Fermi gases \cite{Nascimbene:2010}. We show that $P(\mu,T)$ is directly related to the atom number $N_0=\int n_0(\bs r)\,d^2r$ in our harmonic trap. Indeed, the local density approximation relates the density $n_0(\bs r)$ to the density of the homogenous gas $\ntDh [\mu-V(\bs r),T]$. For an isotropic harmonic potential $V(\bs r)=m\omega^2r^2/2$ the total atom number is
\begin{equation}
N_0=\frac{2\pi}{m\omega^2}\int_{-\infty}^\mu \ntDh(\mu',T)\,d\mu',
\label{eq:pressure1}
\end{equation}
and using the thermodynamic relation $\ntDh=\left(\partial P/\partial \mu\right)_T$, we find $N_0=(2\pi/m\omega^2)P(\mu,T)$. Introducing the dimensionless quantity ${\cal P}=P \ldB^2/\kB T$, which we refer to as the reduced pressure, we then obtain
\begin{equation}
{\cal P}(\mu,T)=\left(\frac{\hbar \omega}{\kB T}\right)^2 N_0 ,
\label{eq:pressure2}
\end{equation}
where $\omega$ is to be replaced by the geometrical mean of $\omega_x$ and $\omega_y$ for an non-isotropic potential.
Our results for the pressure are summarized in Fig.~\ref{Fig:EoS}a, where we plot $\cal{P}$ deduced from Eq. (\ref{eq:pressure2}) as function of $\mu/\kB T$. The temperatures of the data entering in this plot  range from 40~nK to 150~nK. The fact that all data points collapse on the same line show that $\cal{P}$ is a function of the ratio $\mu/\kB T$ only, as expected from  the scale invariance of the system. The HFMF theory is represented by a continuous line in the normal region and by a dotted line in the superfluid region. The dashed line is the Thomas--Fermi prediction at zero temperature ${\cal P}=\pi(\mu/\kB T)^2/\tilde g$. The grey area is the parameter subspace accessible for an ideal Bose gas. Interestingly, although the phase space density ${\cal D}$ can take arbitrarily large values in an ideal 2D Bose gas, one can show that the reduced pressure is $\leq\pi^2/6$, where the equality provides a local criterion for Bose--Einstein condensation in a trapped ideal gas.

We show in Fig.~\ref{Fig:EoS}b our measurements for the phase space density ${\cal D}$, obtained from the central density of each cloud. In wide gray line we plot the prediction of \cite{Prokofev:2002}, which is in good agreement with our results. A measurement of ${\cal D}$ was also reported in \cite{Hung:2011} for a quasi-2D Cesium gas, showing a similar agreement with \cite{Prokofev:2002}. To be quantitative we have fitted the prediction of \cite{Prokofev:2002} to our data by multiplying it by a global factor $\zeta$, and obtained $\zeta=0.93$ as optimal parameter. This $7\%$ discrepancy may be due to residual loss of detectivity in high density regions. Note that the measurement of the pressure \EOS\ is much less sensitive to this possible bias since it relies on the count of the atom number $N_0$ over the whole cloud rather than on the highest value of the spatial density.

From our measurements of ${\cal P}$ and ${\cal D}$ we also obtain the equation of state for the entropy per particle ${\cal S}(\mu, T)$:
\begin{equation}
\frac{\cal S}{\kB}=2\frac{\cal P}{\cal D}-\frac{\mu}{\kB T}\ ,
\label{Eq:entropy}
\end{equation}
which can be derived starting from the entropy per unit area $s=\left(\partial P/\partial T\right)_\mu$, assuming the \EOS\ for ${\cal P}$ to be scale invariant \cite{footnote2}. The corresponding result is shown in Fig.~\ref{Fig:EoS}c. As expected, ${\cal S}$ is large in the non-degenerate regime and rapidly decreases around $\mu/\kB T\approx0.15$, where the superfluid transition is expected for our value of $\tilde g$ \cite{Prokofev:2001}. Finally ${\cal S}$ tends to zero in the Thomas--Fermi regime. Our data points with the largest phase-space density ($\mu/\kB T>0.5$) correspond to ${\cal S}=0.06\,(1)\,\kB$ only. Note that since the BKT transition is of infinite order, one does not expect any discontinuous change for ${\cal P}$, ${\cal D}$ or ${\cal S}$ at the superfluid transition for an infinite homogeneous fluid, although the superfluid density jumps suddenly from 0 to $4/\ldB^2$ \cite{Nelson:1977}.

We now turn to the last part of our study, where we illustrate how to measure the various contributions to the energy of our trapped 2D gases: potential energy $\Ep$ in the external trapping potential,  kinetic energy of the particles $\Ek$, and  interaction energy between atoms $\Ei$. We first point out the simple relation $\Ep=\Ek+\Ei$, obtained from virial theorem assuming 2D contact interaction. We can measure $\Ep=\int n_0(\bs r)\,V(\bs r)\, d^2 r$, from an \emph{in situ} image, but we still need to disentangle the contributions of $\Ek$ and $\Ei$ to the total energy. This can be done by abruptly switching off interactions at time $t=0$. Each particle then undergoes a free harmonic motion $\bs r(t) = \cos (\omega t)\, \bs r(0) + \sin (\omega t)\, \bs v(0)/\omega$. The potential energy after a time $t$ following the switching off of the trapping laser is given by
\begin{equation}
\Ep(t)=\Ep(0)  \cos^2 (\omega t)
+\Ek(0)\sin^2 (\omega t) ,
\label{eq:evolE}
\end{equation}
where we used the fact that the correlation $\langle \bs r(0)\cdot \bs v(0)\rangle$  is zero at thermal equilibrium. Thus we can extract $\Ek(0)$ from the time evolution of $\Ep$, which we obtain from the density profiles at different times $t$.

\begin{figure}[tbp]
\begin{center}
\includegraphics{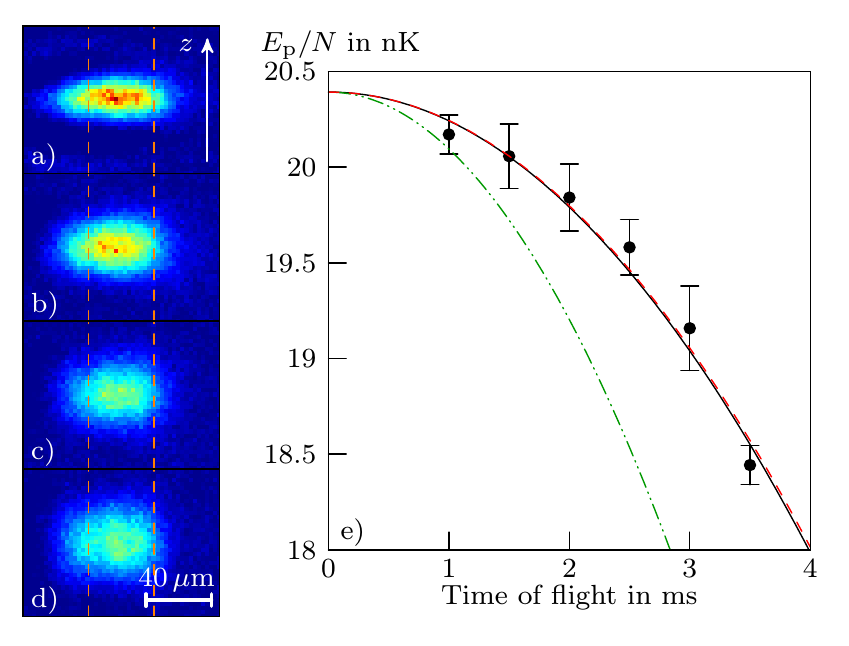}
\end{center}
\vspace{-0.5cm}
\caption{(Color on line) (a) to (d) Side view of a cloud initially in the 2D regime and expanding along $z$ once the laser providing the confinement in this direction has been switched off. (a) $t= 1$\,ms; (b) $t= 2$\,ms; (c) $t= 3$\,ms; (d) $t= 4$\,ms. (e) Time evolution of the potential energy $\Ep$. The different lines represent a fit to the data of a parabola (solid black line), the time evolution assuming completely frozen fluctuations (dashed red line) and the one expected for a dilute non-condensed gas (dash-dotted green line).}
\label{fig:1DTOF}
\end{figure}

In order to implement this procedure, we perform a ``one-dimensional" \TOF\ by switching off abruptly the laser providing the confinement along $z$ while keeping the magnetic confinement in the $xy$ plane. The gas then expands very fast along the initially strongly confined direction $z$, as shown in figures \ref{fig:1DTOF}a to \ref{fig:1DTOF}d, and interactions between particles drop to a negligible value after a time of a few $\omega_z^{-1}$, where $\omega_z^{-1}\sim 100\,\mu$s. The subsequent evolution in the $xy$ plane occurs on a longer time scale given by $\omega^{-1}\sim  8$\,ms. From Eq. (\ref{eq:evolE}) and $\Ek(0)< \Ep(0)$, we expect the size of the gas to decrease for $t\lesssim \omega^{-1}$, which can be understood in simple physical terms. The equilibrium state of the 2D gas results from a balance between the trapping potential, which tends to compress the gas, and the kinetic and interaction energies, which tend to increase its area. When interaction energy drops to zero the equilibrium is broken and the gas implodes in the $xy$ plane. A similar 1D \TOF\ technique was used recently in Boulder with the value of $t$ fixed at $\pi/2\omega$ \cite{Tung:2010}. For this particular choice the initial momentum distribution is converted into position distribution and can thus be measured accurately \cite{Shvarchuck:2002}.

We show in figure \ref{fig:1DTOF}e an example of measurement of $\Ep(t)$ for a gas with $N_0=6.1\,10^4$, $T=72$\,nK, and $\mu/\kB T=0.59$. From the contraction of the gas, we infer $\Ek/\Ep=0.56\,(3)$, from which we deduce $\Ei/\Ep=0.44\,(3)$ using virial theorem. This configuration is thus neither completely in the very dilute regime ($\Ei \ll \Ek\sim \Ep$) nor in the Thomas--Fermi regime ($\Ek \ll \Ei\sim \Ep$) and contains comparable thermal and quasi-coherent fractions.

The measurement of $\Ei$ is of particular interest in this case since it gives access to the density fluctuations in the gas. Indeed, by definition $\Ei=(\hbar^2\tilde g /2m)\int \langle n_0^2(\bs r)\rangle\;d^2r = (\hbar^2\tilde g /2m){\cal F}\int \langle n_0(\bs r)\rangle^2\;d^2r$, where we have introduced the parameter ${\cal F}$ which characterizes the degree to which density fluctuation are reduced. In the limiting case of a very dilute, non-condensed gas, one expects ${\cal F}=2$, since $\langle n_0^2\rangle=2\langle n_0\rangle^2$, while in the opposite limit of suppressed density fluctuations ${\cal F}$=1. Since our measurement provides us with $\Ei$, we can infer the value of ${\cal F}$, from the comparison with the quantity $(\hbar^2\tilde g/2m)\int \langle n_0(\bs r)\rangle^2\;d^2r$, calculated using the \emph{in situ} density profile $n_0$. For the experimental conditions of figure \ref{fig:1DTOF}e, we find ${\cal F}=1.1\,(1)$, very close to 1 that would correspond to completely frozen density fluctuations. Note that this is obtained for a gas still far from  the Thomas--Fermi limit since $\Ek\sim\Ei$. This ``early" freezing of density fluctuations is an important ingredient for the proper operation of the BKT mechanism. This presuperfluid phase, whose existence was also inferred by different methods in \cite{Tung:2010} and \cite{Hung:2011}, constitutes a medium that can support vortices, which pair at the superfluid threshold.

In conclusion we have presented in this Letter various aspects of the thermodynamics of a 2D Bose gas, investigating first the \EOS's for the pressure, the phase space density and the entropy. Our results confirm the scale invariance that was discussed theoretically in \cite{Prokofev:2002} and observed in \cite{Hung:2011} for ${\cal D}$. We point out that the entropy per particle drops notably below $ 0.1\,\kB$ beyond the transition point. With such a low entropy these 2D Bose gases can constitue excellent coolants for other quantum fluids such as 2D Fermi gases \cite{Frohlich:2011}. We have also presented a method that allows one to extract the various contributions to the total energy of the system. By applying it to a degenerate but not fully coherent 2D cloud, we find that density fluctuation are nearly frozen, marking the presuperfluid phase.

\begin{acknowledgments}
We thank Gordon Baym, Yvan Castin, Markus Holzmann, Werner Krauth, Christophe Salomon, Sandro Stringari and Martin Zwierlein for helpful discussions. We are grateful to Benno Rem for his help at an early stage of this project. We are indebted to Aviv Keshet for letting us use the computer code that he wrote for experimental control. This work is supported by IFRAF and ANR (project
  BOFL).
 \end{acknowledgments}


\begin{thebibliography}{22}
\expandafter\ifx\csname natexlab\endcsname\relax\def\natexlab#1{#1}\fi
\expandafter\ifx\csname bibnamefont\endcsname\relax
  \def\bibnamefont#1{#1}\fi
\expandafter\ifx\csname bibfnamefont\endcsname\relax
  \def\bibfnamefont#1{#1}\fi
\expandafter\ifx\csname citenamefont\endcsname\relax
  \def\citenamefont#1{#1}\fi
\expandafter\ifx\csname url\endcsname\relax
  \def\url#1{\texttt{#1}}\fi
\expandafter\ifx\csname urlprefix\endcsname\relax\def\urlprefix{URL }\fi
\providecommand{\bibinfo}[2]{#2}
\providecommand{\eprint}[2][]{\url{#2}}

\bibitem[{\citenamefont{Ho and Zhou}(2009)}]{Ho:2009}
\bibinfo{author}{\bibfnamefont{T.-L.} \bibnamefont{Ho}} \bibnamefont{and}
  \bibinfo{author}{\bibfnamefont{Q.}~\bibnamefont{Zhou}},
  \bibinfo{journal}{Nature Physics} \textbf{\bibinfo{volume}{6}},
  \bibinfo{pages}{131} (\bibinfo{year}{2009}).

\bibitem[{\citenamefont{Mermin and Wagner}(1966)}]{Mermin:1966}
\bibinfo{author}{\bibfnamefont{N.~D.} \bibnamefont{Mermin}} \bibnamefont{and}
  \bibinfo{author}{\bibfnamefont{H.}~\bibnamefont{Wagner}},
  \bibinfo{journal}{Phys. Rev. Lett.} \textbf{\bibinfo{volume}{17}},
  \bibinfo{pages}{1133} (\bibinfo{year}{1966}).

\bibitem[{\citenamefont{Hohenberg}(1967)}]{Hohenberg:1967}
\bibinfo{author}{\bibfnamefont{P.~C.} \bibnamefont{Hohenberg}},
  \bibinfo{journal}{Phys. Rev.} \textbf{\bibinfo{volume}{158}},
  \bibinfo{pages}{383} (\bibinfo{year}{1967}).

\bibitem[{\citenamefont{Berezinskii}(1971)}]{Berezinskii:1971}
\bibinfo{author}{\bibfnamefont{V.~L.} \bibnamefont{Berezinskii}},
  \bibinfo{journal}{Soviet Physics JETP} \textbf{\bibinfo{volume}{34}},
  \bibinfo{pages}{610} (\bibinfo{year}{1971}).

\bibitem[{\citenamefont{{K}osterlitz and {T}houless}(1973)}]{Kosterlitz:1973}
\bibinfo{author}{\bibfnamefont{J.~M.} \bibnamefont{{K}osterlitz}}
  \bibnamefont{and} \bibinfo{author}{\bibfnamefont{D.~J.}
  \bibnamefont{{T}houless}}, \bibinfo{journal}{J. Phys. C: Solid State Physics}
  \textbf{\bibinfo{volume}{6}}, \bibinfo{pages}{1181} (\bibinfo{year}{1973}).

\bibitem[{\citenamefont{Prokof'ev and Svistunov}(2002)}]{Prokofev:2002}
\bibinfo{author}{\bibfnamefont{N.~V.} \bibnamefont{Prokof'ev}}
  \bibnamefont{and} \bibinfo{author}{\bibfnamefont{B.~V.}
  \bibnamefont{Svistunov}}, \bibinfo{journal}{Phys. Rev. A}
  \textbf{\bibinfo{volume}{66}}, \bibinfo{pages}{043608}
  (\bibinfo{year}{2002}).

\bibitem[{\citenamefont{Petrov et~al.}(2000)\citenamefont{Petrov, Holzmann, and
  Shlyapnikov}}]{Petrov:2000a}
\bibinfo{author}{\bibfnamefont{D.~S.} \bibnamefont{Petrov}},
  \bibinfo{author}{\bibfnamefont{M.}~\bibnamefont{Holzmann}}, \bibnamefont{and}
  \bibinfo{author}{\bibfnamefont{G.~V.} \bibnamefont{Shlyapnikov}},
  \bibinfo{journal}{Phys. Rev. Lett.} \textbf{\bibinfo{volume}{84}},
  \bibinfo{pages}{2551} (\bibinfo{year}{2000}).

\bibitem[{\citenamefont{Hadzibabic et~al.}(2006)\citenamefont{Hadzibabic,
  Kr{{\"u}}ger, Cheneau, Battelier, and Dalibard}}]{Hadzibabic:2006}
\bibinfo{author}{\bibfnamefont{Z.}~\bibnamefont{Hadzibabic}},
  \bibinfo{author}{\bibfnamefont{P.}~\bibnamefont{Kr{{\"u}}ger}},
  \bibinfo{author}{\bibfnamefont{M.}~\bibnamefont{Cheneau}},
  \bibinfo{author}{\bibfnamefont{B.}~\bibnamefont{Battelier}},
  \bibnamefont{and} \bibinfo{author}{\bibfnamefont{J.}~\bibnamefont{Dalibard}},
  \bibinfo{journal}{Nature} \textbf{\bibinfo{volume}{441}},
  \bibinfo{pages}{1118} (\bibinfo{year}{2006}).

\bibitem[{\citenamefont{Clad\'e et~al.}(2009)\citenamefont{Clad\'e, Ryu,
  Ramanathan, Helmerson, and Phillips}}]{Clade:2009}
\bibinfo{author}{\bibfnamefont{P.}~\bibnamefont{Clad\'e}},
  \bibinfo{author}{\bibfnamefont{C.}~\bibnamefont{Ryu}},
  \bibinfo{author}{\bibfnamefont{A.}~\bibnamefont{Ramanathan}},
  \bibinfo{author}{\bibfnamefont{K.}~\bibnamefont{Helmerson}},
  \bibnamefont{and} \bibinfo{author}{\bibfnamefont{W.~D.}
  \bibnamefont{Phillips}}, \bibinfo{journal}{Phys. Rev. Lett.}
  \textbf{\bibinfo{volume}{102}}, \bibinfo{eid}{170401} (\bibinfo{year}{2009}).

\bibitem[{\citenamefont{Tung et~al.}(2010)\citenamefont{Tung, Lamporesi,
  Lobser, Xia, and Cornell}}]{Tung:2010}
\bibinfo{author}{\bibfnamefont{S.}~\bibnamefont{Tung}},
  \bibinfo{author}{\bibfnamefont{G.}~\bibnamefont{Lamporesi}},
  \bibinfo{author}{\bibfnamefont{D.}~\bibnamefont{Lobser}},
  \bibinfo{author}{\bibfnamefont{L.}~\bibnamefont{Xia}}, \bibnamefont{and}
  \bibinfo{author}{\bibfnamefont{E.~A.} \bibnamefont{Cornell}},
  \bibinfo{journal}{Phys. Rev. Lett.} \textbf{\bibinfo{volume}{105}},
  \bibinfo{pages}{230408} (\bibinfo{year}{2010}).

\bibitem[{\citenamefont{Hung et~al.}(2011)\citenamefont{Hung, Zhang, Gemelke,
  and Chin}}]{Hung:2011}
\bibinfo{author}{\bibfnamefont{C.-L.} \bibnamefont{Hung}},
  \bibinfo{author}{\bibfnamefont{X.}~\bibnamefont{Zhang}},
  \bibinfo{author}{\bibfnamefont{N.}~\bibnamefont{Gemelke}}, \bibnamefont{and}
  \bibinfo{author}{\bibfnamefont{C.}~\bibnamefont{Chin}},
  \bibinfo{journal}{Nature} \textbf{\bibinfo{volume}{470}},
  \bibinfo{pages}{236} (\bibinfo{year}{2011}).

\bibitem[{\citenamefont{Rath et~al.}(2010)\citenamefont{Rath, Yefsah, G\"unter,
  Cheneau, Desbuquois, Holzmann, Krauth, and Dalibard}}]{Rath:2010b}
\bibinfo{author}{\bibfnamefont{S.~P.} \bibnamefont{Rath}},
  \bibinfo{author}{\bibfnamefont{T.}~\bibnamefont{Yefsah}},
  \bibinfo{author}{\bibfnamefont{K.~J.} \bibnamefont{G\"unter}},
  \bibinfo{author}{\bibfnamefont{M.}~\bibnamefont{Cheneau}},
  \bibinfo{author}{\bibfnamefont{R.}~\bibnamefont{Desbuquois}},
  \bibinfo{author}{\bibfnamefont{M.}~\bibnamefont{Holzmann}},
  \bibinfo{author}{\bibfnamefont{W.}~\bibnamefont{Krauth}}, \bibnamefont{and}
  \bibinfo{author}{\bibfnamefont{J.}~\bibnamefont{Dalibard}},
  \bibinfo{journal}{Phys. Rev. A} \textbf{\bibinfo{volume}{82}},
  \bibinfo{pages}{013609} (\bibinfo{year}{2010}).

\bibitem{footnote1}
The residual anisotropy of the trap is $|\omega_x-\omega_y|/\omega<6\%$ where $\omega=(\omega_x\omega_y)^{1/2}$; it plays no significant role in the subsequent analyses. Our procedure to handle slight deviations with respect to harmonicity is described in the supplementary material.

\bibitem[{\citenamefont{Reinaudi et~al.}(2007)\citenamefont{Reinaudi, Lahaye,
  Wang, and Gu\'ery-Odelin}}]{Reinaudi:2007}
\bibinfo{author}{\bibfnamefont{G.}~\bibnamefont{Reinaudi}},
  \bibinfo{author}{\bibfnamefont{T.}~\bibnamefont{Lahaye}},
  \bibinfo{author}{\bibfnamefont{Z.}~\bibnamefont{Wang}}, \bibnamefont{and}
  \bibinfo{author}{\bibfnamefont{D.}~\bibnamefont{Gu\'ery-Odelin}},
  \bibinfo{journal}{Opt. Lett.} \textbf{\bibinfo{volume}{32}},
  \bibinfo{pages}{3143} (\bibinfo{year}{2007}).

\bibitem[{\citenamefont{Hadzibabic et~al.}(2008)\citenamefont{Hadzibabic,
  Kr{\"{u}}ger, Cheneau, Rath, and Dalibard}}]{Hadzibabic:2008}
\bibinfo{author}{\bibfnamefont{Z.}~\bibnamefont{Hadzibabic}},
  \bibinfo{author}{\bibfnamefont{P.}~\bibnamefont{Kr{\"{u}}ger}},
  \bibinfo{author}{\bibfnamefont{M.}~\bibnamefont{Cheneau}},
  \bibinfo{author}{\bibfnamefont{S.~P.} \bibnamefont{Rath}}, \bibnamefont{and}
  \bibinfo{author}{\bibfnamefont{J.}~\bibnamefont{Dalibard}},
  \bibinfo{journal}{New Journal of Physics} \textbf{\bibinfo{volume}{10}},
  \bibinfo{pages}{045006} (\bibinfo{year}{2008}).

\bibitem[{\citenamefont{Bisset et~al.}(2009)\citenamefont{Bisset, Baillie, and
  Blakie}}]{Bisset:2009}
\bibinfo{author}{\bibfnamefont{R.~N.} \bibnamefont{Bisset}},
  \bibinfo{author}{\bibfnamefont{D.}~\bibnamefont{Baillie}}, \bibnamefont{and}
  \bibinfo{author}{\bibfnamefont{P.~B.} \bibnamefont{Blakie}},
  \bibinfo{journal}{Phys. Rev. A} \textbf{\bibinfo{volume}{79}},
  \bibinfo{pages}{013602} (\bibinfo{year}{2009}).

\bibitem[{\citenamefont{Holzmann et~al.}(2010)\citenamefont{Holzmann,
  Chevallier, and Krauth}}]{Holzmann:2010}
\bibinfo{author}{\bibfnamefont{M.}~\bibnamefont{Holzmann}},
  \bibinfo{author}{\bibfnamefont{M.}~\bibnamefont{Chevallier}},
  \bibnamefont{and} \bibinfo{author}{\bibfnamefont{W.}~\bibnamefont{Krauth}},
  \bibinfo{journal}{Phys. Rev. A} \textbf{\bibinfo{volume}{81}},
  \bibinfo{pages}{043622} (\bibinfo{year}{2010}).

\bibitem[{\citenamefont{Holzmann and Krauth}(2008)}]{Holzmann:2008a}
\bibinfo{author}{\bibfnamefont{M.}~\bibnamefont{Holzmann}} \bibnamefont{and}
  \bibinfo{author}{\bibfnamefont{W.}~\bibnamefont{Krauth}},
  \bibinfo{journal}{Phys. Rev. Lett.} \textbf{\bibinfo{volume}{100}},
  \bibinfo{eid}{190402} (\bibinfo{year}{2008}).

\bibitem[{\citenamefont{Nascimb\`ene et~al.}(2010)\citenamefont{Nascimb\`ene,
  Navon, Jiang, Chevy, and Salomon}}]{Nascimbene:2010}
\bibinfo{author}{\bibfnamefont{S.}~\bibnamefont{Nascimb\`ene}},
  \bibinfo{author}{\bibfnamefont{N.}~\bibnamefont{Navon}},
  \bibinfo{author}{\bibfnamefont{K.~J.} \bibnamefont{Jiang}},
  \bibinfo{author}{\bibfnamefont{F.}~\bibnamefont{Chevy}}, \bibnamefont{and}
  \bibinfo{author}{\bibfnamefont{C.}~\bibnamefont{Salomon}},
  \bibinfo{journal}{Nature} \textbf{\bibinfo{volume}{463}},
  \bibinfo{pages}{1057} (\bibinfo{year}{2010}).

\bibitem{footnote2}
A similar method has been used for a 3D Fermi gas at unitarity, Martin Zwierlein, private communication, February 2011.

\bibitem[{\citenamefont{Prokof'ev et~al.}(2001)\citenamefont{Prokof'ev,
  Ruebenacker, and Svistunov}}]{Prokofev:2001}
\bibinfo{author}{\bibfnamefont{N.~V.} \bibnamefont{Prokof'ev}},
  \bibinfo{author}{\bibfnamefont{O.}~\bibnamefont{Ruebenacker}},
  \bibnamefont{and} \bibinfo{author}{\bibfnamefont{B.~V.}
  \bibnamefont{Svistunov}}, \bibinfo{journal}{Phys. Rev. Lett.}
  \textbf{\bibinfo{volume}{87}}, \bibinfo{pages}{270402}
  (\bibinfo{year}{2001}).

\bibitem[{\citenamefont{Nelson and Kosterlitz}(1977)}]{Nelson:1977}
\bibinfo{author}{\bibfnamefont{D.~R.} \bibnamefont{Nelson}} \bibnamefont{and}
  \bibinfo{author}{\bibfnamefont{J.~M.} \bibnamefont{Kosterlitz}},
  \bibinfo{journal}{Phys. Rev. Lett.} \textbf{\bibinfo{volume}{39}},
  \bibinfo{pages}{1201} (\bibinfo{year}{1977}).

\bibitem[{\citenamefont{Shvarchuck et~al.}(2002)\citenamefont{Shvarchuck,
  Buggle, Petrov, Dieckmann, Zielonkowski, Kemmann, Tiecke, von Klitzing,
  Shlyapnikov, and Walraven}}]{Shvarchuck:2002}
\bibinfo{author}{\bibfnamefont{I.}~\bibnamefont{Shvarchuck}},
  \bibinfo{author}{\bibfnamefont{C.}~\bibnamefont{Buggle}},
  \bibinfo{author}{\bibfnamefont{D.~S.} \bibnamefont{Petrov}},
  \bibinfo{author}{\bibfnamefont{K.}~\bibnamefont{Dieckmann}},
  \bibinfo{author}{\bibfnamefont{M.}~\bibnamefont{Zielonkowski}},
  \bibinfo{author}{\bibfnamefont{M.}~\bibnamefont{Kemmann}},
  \bibinfo{author}{\bibfnamefont{T.~G.} \bibnamefont{Tiecke}},
  \bibinfo{author}{\bibfnamefont{W.}~\bibnamefont{von Klitzing}},
  \bibinfo{author}{\bibfnamefont{G.~V.} \bibnamefont{Shlyapnikov}},
  \bibnamefont{and} \bibinfo{author}{\bibfnamefont{J.~T.~M.}
  \bibnamefont{Walraven}}, \bibinfo{journal}{Phys. Rev. Lett.}
  \textbf{\bibinfo{volume}{89}}, \bibinfo{pages}{270404}
  (\bibinfo{year}{2002}).

\bibitem[{\citenamefont{Fr\"ohlich et~al.}(2011)\citenamefont{Fr\"ohlich, Feld,
  Vogt, Koschorreck, Zwerger, and K\"ohl}}]{Frohlich:2011}
\bibinfo{author}{\bibfnamefont{B.}~\bibnamefont{Fr\"ohlich}},
  \bibinfo{author}{\bibfnamefont{M.}~\bibnamefont{Feld}},
  \bibinfo{author}{\bibfnamefont{E.}~\bibnamefont{Vogt}},
  \bibinfo{author}{\bibfnamefont{M.}~\bibnamefont{Koschorreck}},
  \bibinfo{author}{\bibfnamefont{W.}~\bibnamefont{Zwerger}}, \bibnamefont{and}
  \bibinfo{author}{\bibfnamefont{M.}~\bibnamefont{K\"ohl}},
  \bibinfo{journal}{Phys. Rev. Lett.} \textbf{\bibinfo{volume}{106}},
  \bibinfo{pages}{105301} (\bibinfo{year}{2011}).

\end{thebibliography}

\newpage
\null

\centerline{\Large \textbf{Auxiliary material}}

\vskip 5 mm

\paragraph{Imaging a dense 2D atomic cloud.}
The calibration of absorption imaging consists in relating the number of missing photons on a pixel to the number of atoms on this pixel. The interaction between a probe beam and a single atom is characterized by the absorption cross section $\sigma$ defined by the relation $\gamma=\sigma I/(\hbar \omega_{\rm L})$, where $\gamma$ is the photon scattering rate, $I$ the intensity of the beam on the atoms and $\omega_{\rm L}/2\pi$ its frequency. In the case of a monochromatic resonant beam probing a two-level atom
\begin{equation}
\gamma=\frac{\Gamma}{2}\frac{I}{I+\Is}.
\label{eq.gamma}
\end{equation} In the limit where $I\ll\Is$ the absorption cross section is $\sigma_0\equiv\Gamma\hbar \omega_{\rm L}/2\Is$.

In practice one must take into account stray magnetic fields, non-zero linewidth of the probe laser, optical pumping effects, etc. To model this complex situation, we heuristically replace $\Is$ by an effective saturation intensity $\alpha\Is$ and $\Gamma$ by an effective linewidth $\beta\Gamma$. We then write the number of photons $N_{\rm p}$ scattered during an imaging pulse of given duration $\tau$
\begin{equation}
N_{\rm p}\equiv \gamma\tau= \frac{\beta\Gamma}{2}\frac{ I}{I+\alpha\Is}\tau ,
\end{equation}
or equivalently
\begin{equation}
\qquad \sigma=\sigma_0\frac{\beta}{\alpha+I/\Is}.
\label{eq.crossSection}
\end{equation}
At low intensity $N_{\rm p}$ is proportional to $I$ as in the two-level case, but with a multiplicative coefficient $\beta/\alpha$ due (for example) to the broadening of the resonance line. At large intensity the number of scattered photons saturates at $\beta\Gamma \tau /2$ instead of $\Gamma\tau/2$, which models a reduction that can be caused by optical pumping effects,  for instance.

We now turn to the description of absorption imaging of a 2D atomic cloud. The imaging process consists in shining a resonant laser beam on an atomic sample, and in imaging the transmission of the sample on a camera. In order to relate the missing photon number to the atomic density $n(x,y)$, we calculate the probability for a photon of the probe beam to reach a pixel of the camera. We introduce the area $A$ associated to this pixel in the atomic plane. In the limit where $\sigma\ll A$ a photon has a probability $\sigma/A$ to be absorbed by a given atom, hence a probability $P_{\rm t}=(1-\sigma/A)^N$ to be transmitted, where $N$ is the number of atoms in the area $A$. Thus we find:
\begin{equation}
P_{\rm t}\approx e^{-\sigma n},
\label{eq.proba_trans}
\end{equation} where we have used $n=N/A$, assuming that the atomic density varies smoothly over the pixel size. The intensity of the beam at the output of the cloud is $\If=P_{\rm t}\Ii$ and we obtain:
\begin{equation}
-\ln\left(\frac{\If(x,y)}{\Ii}\right)=\sigma\,n(x,y) ,
\label{eq.BeerLambert}
\end{equation}
where $\sigma$ depends on the effective intensity $I$ on the atoms [Eq. (\ref{eq.crossSection})]. If the optical thickness of the cloud is large, \emph{i.e.} if the intensity $\If$ just after the plane of atoms  is significantly lower than the intensity $\Ii$ just before this plane, the effective intensity $I$ must be determined in a self-consistent manner by imposing:
\begin{equation}
\If=\Ii-n\,\sigma(I)\,I.
\label{eq.Ieff}
\end{equation}
The elimination of the effective intensity $I$ from Eqs. (\ref{eq.crossSection})-(\ref{eq.Ieff}) yields:
\begin{equation}
n\sigma_0\,\beta=-\alpha\ln\left(\frac{\If}{\Ii}\right)+\frac{\Ii-\If}{\Is}.
\label{eq.DGO}
\end{equation}
It is interesting to note that even though the derivation in a 2D system differs from the 3D case, the result is similar to the one given in \cite{Reinaudi:2007}. The first member of the right-hand side of Eq. (\ref{eq.DGO}) is dominant in the weak intensity limit, and corresponds to the 2D analog of the 3D Beer--Lambert law. In the high intensity limit, the second member of the right-hand side dominates.
We calibrated $\alpha=2.6\,(3)$ using the same method as in \cite{Reinaudi:2007}: we performed absorption imaging of clouds obtained in similar experimental conditions with various intensities $\Ii$ ranging from $0.1\,\Is$ to $6\,\Is$, and  imposed that these measurements provide the same result for the left-hand side of  Eq. (\ref{eq.DGO}). Here we restricted ourselves to low atomic density regions, to ensure that collective effects in the optical response of the gas were negligible.  The calibration of $\beta=0.40\,(2)$ was performed as in \cite{Rath:2010b}, using the HFMF prediction as a fit to the low-density parts of our atomic distributions, and using $\mu$, $T$ and $\beta$ as optimization parameters. In \cite{Rath:2010b} where only low intensity imaging was used, this calibration provided the detectivity factor $\xi$, which is related to the present parameters $\alpha$ and $\beta$ by $\xi=(15/7)\beta/\alpha$.

\begin{figure}[tbp]
\begin{center}
\includegraphics{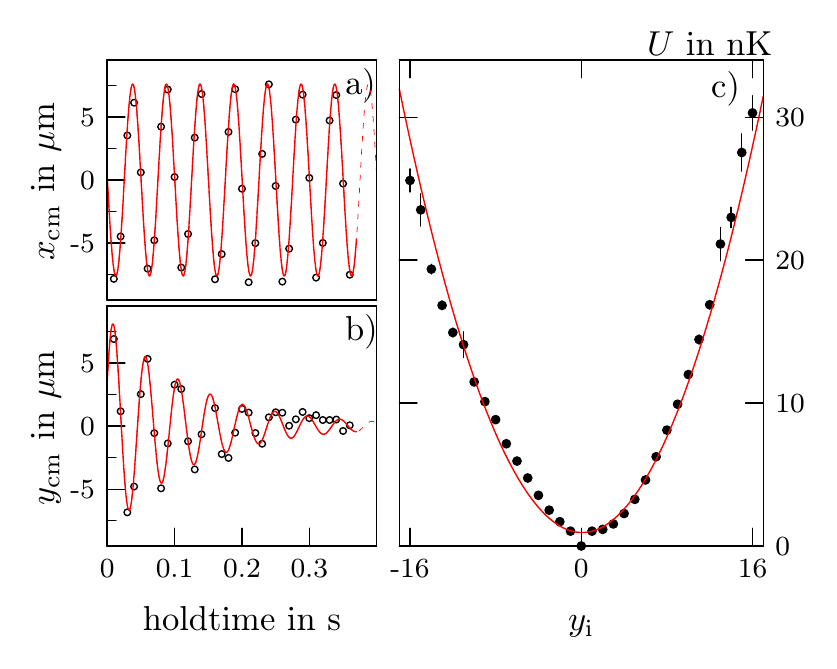}
\end{center}
\vspace{-0.5cm}
\caption{(Color on line) (a) and (b) Center of mass oscillations (hollow circles $\circ$) along $x$ (a) and $y$ (b). The red lines correspond to a fit with a sine (a) and a damped sine (b). (c) Reconstructed potential along the $y$ axis (filled circles $\bullet$) and a harmonic fit (red line).}
\label{Fig:osc_and_pot}
\end{figure}

\paragraph{Obtaining a density profile from an absorption image.}
The confinement potential in the $xy$ plane is essentially provided by our magnetic trap, but it may also be affected by some imperfections in the intensity profile of the beam that freezes the $z$ degree of freedom. These imperfections are revealed by looking at the center of mass oscillations $x_{\mathrm{cm}}(t)$ and $y_{\mathrm{cm}}(t)$, shown in Fig.~\ref{Fig:osc_and_pot}a. Whereas the oscillation along the direction of propagation of the ``freezing laser" ($x$) shows no deviation with respect to harmonic motion, the oscillation along $y$ is damped. This is likely caused by   irregularities of the transverse intensity profile of the freezing laser. In order to cope with these defects we have abandoned the standard technique consisting in making angular average of the images to produce radial density profiles. Instead we take advantage of the separability of the potential in the $xy$ plane: $V(x,y)= m\omega_x^2 x^2/2 + U(y)$, where $U(y)$ accounts for the magnetic trapping potential and the irregularities of the freezing laser. We consider  cuts of the measured density profile along the $x$ direction, measured for various  $y_i$'s with $i=1,\ldots,q$. In practice, we consider the $q=31$ central lines of our images. We expect that two cuts corresponding to $y_1$ and $y_2$ coincide, provided we shift the second one by making the substitution $m\omega_x^2 x^2$ $\to$  $m\omega_x^2 x^2 + U(y_2)-U(y_1)$. In practice we  perform a least-square fit to optimize the superposition of the various cuts,  taking the numbers $U(y_i)$  as parameters. We use a single set of $U(y_j)$ to fit a whole series of images taken at a given temperature. The robustness of the procedure is excellent, as shown in Fig.~\ref{Fig:osc_and_pot}b, where we give the reconstructed potential $U(y)$, with bars corresponding to the statistical errors of the $U(y_j)$'s for various series of images acquired at different temperatures.

\paragraph{Analyzing a density profile.}
For each configuration of a 2D cloud, we first take a low-intensity image and then a high-intensity image in the following run. From the density profile of the low-intensity image, we determine the temperature $T$ and the chemical potential $\mu$ by fitting the low density region with the HFMF prediction. Our fitting function takes into account the residual excitation of the $z$ degree of freedom. The high-intensity image provides the density profile $n(\bs r)$.

\begin{figure}[b]
\begin{center}
\includegraphics{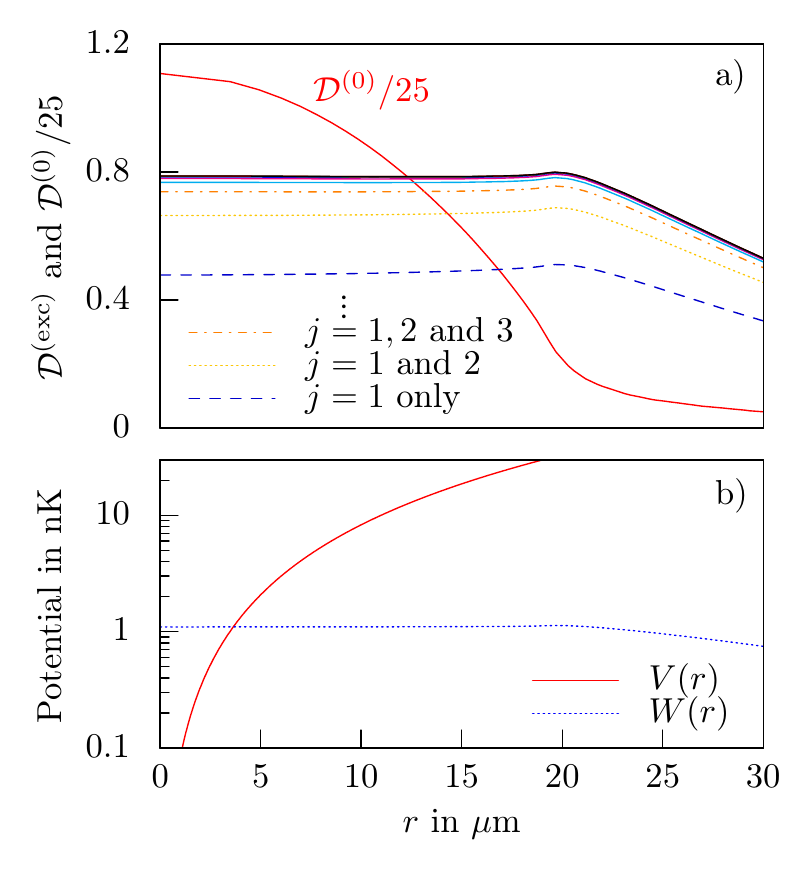}
\end{center}
\vspace{-0.5cm}
\caption{(Color on line) (a) Phase space density of the ground (solid red line) and excited state(s) of the $z$-motion. The $n$-th line from the bottom corresponds to the contributions of excited levels 1 to $n$. (b) Comparison of the trapping potential (red solid line) and the repulsive potential created by the excited atoms on the population in the ground state (blue dotted line).}
\label{Fig:modified_pot}
\end{figure}

Once $T$ and $\mu$ are known, we self-consistently determine the population of the excited states using the method described in \cite{Hadzibabic:2008,Tung:2010}, assuming the atoms in the excited states $j\geq 1$ of the $z$ motion to be in the HFMF regime. In practice we restrict the analysis to the first ten levels. In order to give an estimation of the contribution of the various levels $j\geq 1$ to the total density, we show in Fig.~\ref{Fig:modified_pot}a numerical results obtained by applying this procedure to a numerically generated profile, produced using the prediction \cite{Prokofev:2002} with $T=100\,$nK and $\mu/\kB T = 0.45$. This temperature is on the high side of our experimental range, where the influence of the atoms in the excited states along $z$ is expected to be the most important. We plot in Fig.~\ref{Fig:modified_pot}a the phase space density of the excited states ${\cal D}^{(\mathrm{exc})}$, distinguishing the contribution of the state(s) $j=1$, $j=(1,2)$, $j=(1,2,3)$, etc. For comparison we also plot the profile ${\cal D}^{(0)}$ obtained from \cite{Prokofev:2002}, associated to the atoms in the ground state. Note that the contribution of the states $j>4$ is already negligible. The phase space density associated to each excited state is lower than 0.5, which justifies to treat the atoms in these states within the HFMF approximation. The flattened shape of the density distributions in the central region is due to the repulsive interaction with the atoms in the ground state of the $z$ motion.

This procedure also allows us to calculate the effective potential felt by the atoms in $j=0$, when the repulsive potential $W(\bs r)$ created by the atoms in $j\geq 1$ is taken into account. Plotting together $W(\bs r)$ and the trapping potential $V(\bs r)$ (Fig.~\ref{Fig:modified_pot}b) we see that $W(\bs r)$ is essentially negligible ($\lesssim 1$\,nK) and  one can thus consider the density $n_0(\bs r)$ to be insensitive to the presence of the atoms in $j\geq 1$.

\paragraph{Effect of the finite interaction energy.}
The expression $\tilde g=\sqrt{8\pi}a/\ell_z$ for the interaction strength assumes that the atoms are confined in the gaussian single-particle ground state of the harmonic motion along the $z$ direction. However, because  the interaction energy is not completely negligible compared to $\hbar \omega_z$, the state of the $z$ motion at $T=0$ is modified by these interactions, which in turn modifies the coupling strength $\tilde g$. To estimate the corresponding effect we used first-order perturbation theory to determine the modified ground state $\varphi_0(z)$ of the $z$ motion. We then calculated the coupling strength $\tilde g$, which is proportional to $\int |\varphi_0|^4\,dz$. We find that $\tilde g$ is reduced by the factor $\approx 1-1.4\, \ntD a \ell_z$, which is $\approx 0.9$ at the center of our densest clouds, comparable to the noise level on our data.

\end{document}